\documentclass[twocolumn,showpacs,preprintnumbers,amsmath,amssymb,prl]{revtex4}
\usepackage{graphicx}
\usepackage{bm}
\usepackage{syntonly}
\def\be{\begin{equation}} 
\def\ee{\end{equation}}
\def \beal{\begin{align}}
\def \eeal{\end{align}}
\def\bea{\begin{eqnarray}} 
\def\eea{\end{eqnarray}}
\def \line{\hbox to \hsize}    
\def\frac #1#2{{#1\over #2}}

\def\psid{\psi^{\dagger}}

\def\1{\mbox{\bf 1}}
\newcommand{\comment}[1]{}

\begin{document}

\title{Collective modes and electromagnetic response of a chiral superconductor}

\author{Rahul Roy and Catherine Kallin}

\affiliation{ Department of Physics and Astronomy, McMaster University\\Hamilton, Ontario, Canada L8S 4M1}   

\begin{abstract}
 Motivated by the recent controversy surrounding the Kerr effect measurements in strontium ruthenate \cite{xia:167002},
we examine the electromagnetic response of a clean chiral p-wave superconductor. When the contributions of 
the collective modes are accounted for, the Hall response in a clean chiral superconductor is smaller by 
several orders of magnitude than previous theoretical predictions and is too small to explain the experiment. 
We also uncover some unusual features of the collective modes
of a chiral superconductor, namely, that they are not purely longitudinal and couple to external 
transverse fields.
\end{abstract}

\maketitle

Recent optical experiments by Xia {\it et al.} \cite{xia:167002} on the polar Kerr effect in 
strontium ruthenate, Sr$_2$RuO$_4$, have been interpreted as evidence for broken time reversal
symmetry in the superconducting state. Early indications of broken time reversal
symmetry in superconducting Sr$_2$RuO$_4$ came from muon spin resonance experiments \cite{luke1998trs}.
These results, together with crystal symmetry and energetic considerations \cite{sigrist-review}, 
point to a chiral p-wave superconducting order analogous to the superfluid order of the $^3$He-A phase \cite{mackenzie,leggett}. 
Recent Josephson interferometry measurements \cite{kidwingira2006dso} have also been interpreted
as evidence for chiral p-wave order in superconducting Sr$_2$RuO$_4$. Such a superconducting state, 
if confirmed, is expected to have many exciting implications for exotic physics \cite{dassarma}.    

The experiment by Xia {\it et al.} found a Kerr angle of approximately 60 nanoradians at a frequency, $ \omega \simeq 0.8 $\,eV,
which is 
small compared to the Kerr angle observed in typical ferromagnets, but can be understood qualitatively
if one notes that the superconducting gap (or order parameter) for Sr$_2$RuO$_4$ is substantially
reduced from that of a typical ferromagnet \cite{xia:167002}.  
The Kerr angle at high frequencies is related to the ac Hall conductivity. Theoretical work, however, 
is divided on the issue of whether the experimental observation 
result is consistent with the linear response theory of a clean chiral p-wave superconductor \cite{mineev, yakovenko, yip}. 
Earlier works on the electromagnetic response of a chiral p-wave superconductor, with an 
isotropic Fermi surface \cite{yip,joynt}, found no quasi-particle contribution in the clean limit, even 
in the presence of particle-hole asymmetry, but did predict a Kerr angle due to the so-called 
``flapping'' collective mode.  The predicted magnitude is smaller by several orders of 
magnitude than that observed in Sr$_2$RuO$_4$.  

These earlier theoretical studies neglected the effect of an anomalous density-current correlation function 
which vanishes for a non-chiral 
superconductor.  
It was recently argued that when this anomalous correlation function is taken into account, the linear response theory does predict an ac Hall conductivity which is large enough to account for the experiments \cite{yakovenko, mineev}. In the effective action language, the anomalous correlation function gives rise to a 
Chern-Simons-like term which is reminiscent of the quantum Hall effect \cite{volovik1}. 
A similar term in the Ginzburg Landau free energy leads to a small ``spontaneous Hall effect'' in 
a finite system \cite{furusaki}. 
When the dynamics of the spin degrees of freedom which are unimportant for the purposes of the present 
paper are also considered, the effective action of a chiral superconductor or superfluid also contains  a non-abelian Chern Simons term which is responsible for the spin quantum Hall effect in these systems \cite{volovik1989fcs,senthil,read,sengupta2006she}. 

In this paper, we examine afresh the linear response of a chiral p-wave superconductor taking into 
account the anomalous density-current correlation functions and also the contributions from collective 
modes.  We find that when both factors are taken into account, the ac Hall response is strictly zero 
for an idealized beam normally incident on the a-b plane, with no in-plane wave vector. This is
in contrast to the recent results which also considered the effects of the anomalous correlation function \cite{yakovenko,mineev}. As will
be shown below, the discrepancy can be attributed to the contribution from a term in the supercurrent 
response which was assumed to be negligible at high frequencies \cite{yakovenko} but which, in fact, exactly
cancels the effect found for zero in-plane wave vector. 
Nevertheless, for EM waves with a small non-zero in plane wave vector, the ac Hall conductivity is 
small but non-zero as has been previously shown in two-dimensions \cite{goryo}.  A byproduct of our 
investigations is the uncovering of some unusual features of the collective modes such as  a novel 
coupling between the collective modes and transverse EM waves and a transverse current associated 
with these modes.    

A chiral superconductor has many properties that are markedly different from a non-chiral superconductor. 
The chirality of a $p+ip$ superfluid leads to the presence of edge states which in turn give rise to a 
macroscopic edge current in a neutral superfluid \cite{kita,stone}. In a superconductor, these edge currents are screened 
due to the Meissner effect; nevertheless, they are predicted to be substantial. In both cases, the 
presence of these edge currents can be traced to the presence of the Chern-Simons-like term in the 
effective action and in the Ginzburg Landau expression for the free energy \cite{furusaki, stone}. 
These edge currents in a neutral chiral superfluid, such as a droplet of superfluid $^3$He in the  
A phase, contribute a macroscopic angular momentum which is proportional to the number of particles 
in the system \cite{mermin,kita}.
The failure of experiments to detect these edge states \cite{kirtley,moler} as well as 
other signatures of chirality in strontium ruthenate has led to doubts about the validity of the 
proposed chiral order parameter structure \cite{mackenzie}.  Therefore, it is of considerable interest to determine
whether the optical Kerr measurements on strontium ruthenate can be explained within the theory of
chiral p-wave superconductivity.

We begin by using the low energy effective action of a two dimensional chiral superconductor to 
study its linear response in Section I.  Using the continuity equation, we show that, in contrast 
to a non-chiral superconductor, the collective modes are 
excited by a transverse electromagnetic wave.  We then study linear response in a three dimensional 
theory in Sec. II and obtain expressions
for the Hall conductivity  in Sec. III.  Finally, we discuss the polar Kerr effect and compare 
our calculations with experiments in section IV.

 \section{I. Linear Response and Effective Action in a 2D Model} 

The linear response of a planar chiral superconductor can be studied using the low energy phase only 
effective action which is obtained by integrating out the fermionic degrees of freedom from the 
theory :   
\bea
 e^{i S_{\rm{eff}}} = \int d[\psi]d[\psi^{\dagger}] \exp({i\over2}\int d^2 x dt \left(\psid (\partial_t - H) \psi\right)), 
\eea
 where $\psi, \psid$ are the Grassmanian fields in the two component Nambu formalism and H is the Bogoliubov 
de Gennes Hamiltonian.  For a 2D chiral superconductor,   this effective 
action can be written as follows \cite{goryo, goryo1,stone}: 
\begin{align}
& S_{\rm eff}(A,\Phi)= - \int d^2xdt \left[\rho_s\left(\frac{\partial \Phi/2}{\partial t} +eA_0\right) \right.\nonumber\\
& +\frac {\rho_s}{2m}\left\{\left(\bm{\nabla}\Phi/2-e{\bf A}\right)^2  -\frac{1}{c_s ^2}\left(\frac{\partial \Phi/2}{\partial t}
+e A_0\right)^2 \right\}\nonumber\\
& \left. +c_{xy} \left(\frac{1}{e}\frac{\partial \Phi/2}{\partial t}
+A_0\right)\left(\bm{\nabla}\times {\bf A}\right)_z \right], \label{2dea}
\end{align}
where $\rho_s $ is the equilibrium superfluid density, $\Phi({\bf r},t)$ is the phase of the 
superconducting order parameter and  $c_s$ is the speed of sound.
The first two terms are the only terms present for an s-wave superconductor in the London limit of
constant superfluid density, $\rho_s$, and the third term arises from fluctuations of the superfluid density. 
The above effective action is applicable for a single sheet of a chiral superconductor. The effective 
action for n decoupled sheets is obtained by multiplying the above action by n. 
The term containing $c_{xy}$ is the Chern-Simons-like term which, as pointed out above, arises 
from the density-current correlation function. In general, $c_{xy}$ is frequency dependent as
discussed below. The $c_{xy}$ term is not the full Chern 
Simons term, because it is missing the term $ \epsilon_{0ij}\bm{A}_i\partial_t \bm{A}_j $ \cite{stone,goryo}.
 The action is nevertheless invariant under gauge transformations:
$\Phi\to \Phi+2\theta, {\bf A} \to {\bf A}+ {1\over e}\bm{\nabla}\theta,
 A_0 \to A_0 - {1\over e}\partial_t \theta$.

The current density response is obtained in the usual manner:
\bea
\bm{j} = \frac{\delta S}{\delta \bm{A}} = \frac{e\rho_s}{m}(\bm{\nabla}\Phi/2 - e \bm{A} ) +
\bm{j}^{\rm {cs}},
\eea
where the anomalous part of the current density, arising from the Chern-Simons-like term, is
 \bea
 \bm{j}^{\rm{cs}} =  c_{xy} \bm{\hat{z} \times }\left[  \bm{\nabla} A_0  +
 \frac{\partial ( \bm{\nabla}\Phi /2) } {e\partial t} \right].
 \eea
Similarly, the charge density is:
\be
\rho= - \frac{\delta S_{\rm eff}}{\delta A_0}= e\rho_s - \frac
{e\rho_s}{mc_s^2}\left(\frac{\partial \Phi/2}{\partial t}
+ e A_0\right)+ c_{xy} (\bm{\nabla}\times {\bf A})_z. \label{3}
\ee

 We note that the anomalous current density can be rewritten in the following form: 
 \bea
   \bm{j}^{\rm{cs}} = c_{xy} \bm{\hat{z} \times }\left[ -\bm{E} +  
\frac{\partial ({\bm{\nabla}\Phi/2 - e \bm{A})}}{e\partial t}\right].    \label{jcs1}
 \eea
At high frequencies, in the long wavelength limit,  the expressions for the anomalous current response is 
obtained by simply replacing the coefficient $c_{xy}$ by a frequency dependent one $c_{xy}(\omega)$ as 
obtained from the form of the correlation function.  

It was argued that at high frequencies, the second term in Eq. (\ref{jcs1}) can be neglected, giving rise to an 
anomalous current which flows perpendicular to the electric 
field \cite{yakovenko,mineev}.  The expression for the Hall conductivity thus obtained 
agrees quite well with the value extracted from the observed optical Kerr effect.  However, in a non-chiral
superconductor, the second term is proportional to the time-derivative of the superfluid current, which itself is 
proportional to the electric field, $\partial {\bf j}/\partial t = \rho_s e^2 {\bf E}/m$, so that 
the two terms exactly cancel, giving a vanishing Hall conductivity.  This is also the case
for a clean chiral superconductor, since this relation between the current and the electric field
follows from translational symmetry, and must hold independent of any interactions which lead to
superconductivity.  In the clean limit, the ac conductivity is simply
$\sigma_{\alpha,\beta}(k=0,\omega)=\delta_{\alpha,\beta}\rho_s e^2/(m\omega+i\epsilon)$, so that
the bulk ac Hall conductivity, $\sigma_{xy}(k=0,\omega)$, vanishes at all frequencies \cite{read}.  It is
possible to have a small spontaneous dc Hall conductivity in a finite sample, due to edge effects \cite{furusaki}. 
Furthermore, both the Chern-Simons-like term, and contributions from the flapping collective modes can give rise to a spontaneous 
ac Hall response at finite wavevector, even in the ideal, clean limit \cite{yip,goryo,golub}.

The finite wave vector response can be studied by integrating out the superconducting
phase from the effective action, to obtain an expression for the current response as a 
function of the total field \cite{goryo,golub}.
Here, we adopt a quantum hydrodynamic approach by starting from the above effective action and using the 
continuity equation to determine the dynamics of the collective modes.
This approach makes the contributions from the collective modes and their coupling to the 
electromagnetic waves transparent.  Approaches which are somewhat similar have also appeared in the 
literature \cite{arseev,ambegaokar}.

The equation of motion for the phase of the order parameter is simply the continuity equation: 
     \bea 
  \bm{\nabla\cdot j} + \frac{\partial \rho}{\partial t}= 0 
  \eea
In this section, we consider a two-dimensional model (with no dependence on z, so that in Fourier
space, $k_z$=0) and work in the transverse gauge, 
${\bf\nabla}\cdot{\bf A}$=0, for simplicity. In this case, the 
above continuity equation yields:  
    \bea
     \frac{e\rho_s}{m}\bm{\nabla}^2 ( \Phi/2) - \frac{e\rho_s}{m c_s ^2} \left(\frac{\partial^2  
\Phi/2}{\partial t^2} - e \frac{\partial A_0}{\partial t} \right)
    + c_{xy} \frac{\partial B_z}{\partial t}  = 0.      \label{coll3}
    \eea
It follows that, in the transverse gauge, the phase is decoupled from the field except when at least one 
of $\frac{\partial B_z}{\partial t} $ or $ \frac{\partial A_0}{\partial t} $ is nonzero. 
The coupling of the phase variable to a transverse field is a novel 
feature of a chiral superconductor.
 Unlike a conventional superconductor where such a coupling may arise 
due to mass anisotropy \cite{millis}, in a chiral superconductor, 
this coupling persists even when the Hamiltonian has Galilean invariance \footnote{More generally, unlike in a non-chiral superconductor, this coupling persists when the Hamiltonian has a three fold or higher rotation symmetry.}. Due to this coupling, an external  
transverse EM wave can give rise to charge density oscillations which generate a scalar Coulomb potential $A_0$.  The dispersion 
of these modes will be examined in the next section.  
  There is an appealing physical picture for this coupling. It was noted in Ref. \cite{stone} that the  coupling of the density to the z-component of 
the magnetic field in Eq. (\ref{3}) can be understood as arising from the diamagnetic coupling of the Cooper pairs which have an intrinsic magnetic 
moment (due to the nonzero angular momentum), with the external magnetic field. In this case, when the magnetic field oscillates in time, this coupling 
gives rise to density oscillations and hence excites the collective modes. In a charged superfluid, these density oscillations give rise to an internal field described by the scalar Coulomb potential in Eq. (\ref{coll3}). 
 
In linear response, one can simply take the Fourier transform of the above equation to obtain 
 \bea
 \frac{e\rho_s }{mc_s ^2}(\omega^2 -c_s ^2 k^2) \Phi(\bm{k},\omega)/2 =  i c_{xy} \omega B_z  
 -{e^2\rho_s \over m c_s ^2}i\omega A_0 ,
  \label{coll1}
 \eea
  where ${\bf k}$ is a 2d vector.  Thus the collective phase field excited by the magnetic field is given by
 \bea
 \Phi(\bm{k},\omega) /2=  \frac{i\omega \left(m c_s ^2 c_{xy}B_z - e^2\rho_s A_0 \right)}
{e\rho_s (\omega^2 - c_s ^2 k^2)}.
\label{coll2}
 \eea
Putting this back into the equation for the current response, we see that the collective mode contributes in  \textit{two} ways.
 It enters into the usual term, as well as into the anomalous or the Chern-Simons-like term. 

 We write the current as a sum of two terms, $\bm{j} = \bm{j}_d + \bm{j}_{\Phi}$, where $\bm{j}_d$ is the 
``direct" response and is given by 
\bea
\bm{j}_d = \frac{-e^2\rho_s}{m}( \bm{A} ) + c_{xy}(\bm{\hat{z}}\times i\bm{k}A_0).
\eea
This is the only term which exists when the collective modes are not excited, i.e., when 
$dB_z / dt = 0$ and $dA_0/dt =0$, and is the only term to contribute at $k=0$.  The other term, $\bm{j}_{\Phi}$, 
is the current response due to the collective modes and is given by
\bea
\bm{j}_{\Phi} &=& \frac{e\rho_s}{m}(\bm{\nabla}\Phi/2  ) +
c_{xy} \bm{\hat{z} \times } \frac{\partial ( \bm{\nabla}\Phi /2) } {e\partial t} \\
  & =& \left[({e \rho_s \over m}) i {\bm{k}} + {c_{xy}\omega \over e} \hat{\bm{z}} \times {\bm{k}}\right]
   \Phi(\bm{k},\omega) /2 . \label{1}
 \eea

 The Hall current $\bm{j}_{H}$, comes from terms in the current response  that are linear in $c_{xy}$ :
\bea
   \bm{j}_H =  \frac{c_{xy} c_s^2}{(\omega^2 - c_s^2k^2)}   \left( k^2\hat{\bm{z}} \times ( - i\bm{k} A_0)
-  \bm{k}\omega B_z \right) . 
\eea

    Rewriting $ -i\bm{k}A_0$ as $ \bm{E} - i \omega \bm{A} $ and noting that 
$ B_z = -i (\bm{\hat{z}}\times \bm{A})\cdot\bm{k}$ and  $\bm{k}\cdot\bm{A}=0$, the 
expression for the Hall current reduces to 
\bea
 \bm{j}_{H} =   c_{xy}\frac{c_s ^2 k^2 (\hat{\bm{z}} \times \bm{E})}{\omega^2 - c_s ^2 k^2} .
\eea

Thus the  Hall conductivity, defined in this geometry as $\sigma_{xy}=j_x/E_y $, is:
 \bea
  \sigma_{xy} = c_{xy}\frac{-c_s ^2 k^2}{\omega^2 - c_s ^2 k^2}. \label{4}
 \eea
 
 We have thus recovered the result of Refs. \cite{goryo,golub}  that the Hall conductivity has 
a $ k^2 $ dependence and vanishes in the limit $ k\rightarrow 0 $ at finite $\omega$, as required by 
Galilean invariance. 
 The above analysis was restricted to EM waves propagating in the plane of a 
two dimensional material. 
  A more general analysis is presented in the next section. 
   
  \section{II. 3D layered model and Collective modes} 
In order to connect to experiments on
Sr$_2$RuO$_4$, we need to go beyond the two-dimensional model
considered above.  Sr$_2$RuO$_4$ is extremely
anisotropic, but with coherent transport along
the c-axis \cite{katsufuji}. The low energy effective action describing such a
layered superconductor can be written, in analogy with Eq. (\ref{2dea}), as:

\begin{align}
& S_{\rm eff}(A,\Phi)= - \int d^3xdt \left[\rho_s\left(\frac{\partial \Phi/2}{\partial t} +eA_0\right) \right.\nonumber\\
& +\sum_i\frac {\rho_s}{2m_i}\left(\partial_i\Phi/2-e{A_i}\right)^2  -\frac{\rho_s}{2m c_s ^2}\left(\frac{\partial \Phi/2}{\partial t}
+e A_0\right)^2 \nonumber\\
& \left. + \tilde{c}_{xy\hat{z}} \left(\frac{1}{e}\frac{\partial \Phi/2}{\partial t}
+A_0\right)\left(\bm{\nabla}\times {\bf A}\right)_z \right], \label{3dea}
\end{align}
where $\rho_s$ now represents the superfluid density in three dimensions and the mass parameters $m_i$ have been used 
 to represent the anisotropic diamagnetic current response in the 
long wavelength limit. We restrict ourselves to the case where $m_x = m_y = m $.  
 In the two dimensional limit, $m_z \rightarrow \infty$, the Chern Simons coefficient $c_{xy\hat{z}}$ 
becomes $ c_{xy\hat{z}} = {c_{xy} \over a}$, where $a$ is the 
 interlayer separation.  More generally, we can write $ c_{xy\hat{z}} = \alpha c_{xy}$ where $\alpha $ is a parameter 
 which has the dimensions of inverse length. Hereafter, we use the notation $\tilde{c}_{xy} $. It should also be noted 
 that the velocity of sound, $c_s$, is no longer that applicable for the two dimensional case and 
depends on the details of the microscopic Hamiltonian.  Here, we do not specialize to a particular Hamiltonian.  \\

 \def \cxy {\tilde{c}_{xy}}
  The current density is then given by 
  \bea
  \bm{j} = -\frac{e^2 \rho_s}{m} \bm{\tilde{A}} + \cxy (\bm{\hat{z} \times \bm{\nabla}} A_0) +  \nonumber \\
    \frac{e \rho_s}{m} i \tilde{\bm{k}} (\Phi/2)  + \frac{\cxy \omega}{e} (\hat{\bm{z}} \times \bm{k}) (\Phi/2),
    \label{current}
  \eea
   where $ \tilde{\bm{k}} = ( k_x, k_y, k_z\frac{m}{m_z})$ and
  $\tilde{\bm{A}} = (A_x, A_y, A_z \frac{m}{m_z}) $. 
  The expression for the charge density has the same form as Eq. (\ref{3}) with the parameter $c_{xy}$ replaced by $\cxy$ and $\rho_s$ by the appropriate three dimensional superfluid density.

  The collective mode response can again be determined using the continuity equation, as in the two 
dimensional case,  and gives:
 \bea
  \Phi /2(\bm{k},\omega) = \frac{ \cxy i\omega B_z + {e^2 \rho_s \over m} i \tilde{\bm{k}}\cdot\bm{A} -
   {e^2 \rho_s \over m c_s ^2} i\omega A_0}
  {{e\rho_s \over m c_s^2}(\omega^2 - c_s ^2\bm{k\cdot\tilde{k}})}. \label{5}
 \eea

In calculating the electromagnetic response, the Coulomb interaction does not appear explicitly
because all fields and gauge potentials correspond to the total fields.  However, to obtain
the dispersion relation for the collective modes, one needs to include the effect of Coulomb
interactions.  This is most simply done by using the self consistent field 
implementation of the random phase approximation \cite{ehrenreich}. We replace $A_0 $ by 
$ \frac{4\pi}{\epsilon k^2}$ (where $\epsilon$ is the dielectric constant of the system that comes 
from sources other than the conduction electrons) in Eq. (\ref{current}), set all other fields to zero 
and use the continuity equation.  On doing this we obtain : 
\bea
\left[\frac{e \rho_s \omega^2}{m c_s^2} (1+ \frac{\omega_p^2}{c_s^2 k^2})^{-1} - \frac{e \rho_s}{m} 
\bm{k}\cdot \bm{\tilde{k}}\right] \frac{\Phi}{2} = 0 .
\eea
 This gives the dispersion of the collective modes to be 
\bea
 \omega^2 = c_s^2 \left( 1+ \frac{\omega_p^2}{\epsilon c_s^2k^2} \right) \bm{k}\cdot\bm{\tilde{k}},
\eea
 where $\omega_p ^2 = {4\pi \rho e^2 \over m}$.  The 
 dispersion is identical to that of an anisotropic s-wave superconductor, where the anisotropy enters
through $\bm{\tilde{k}}$. 
 The long-wavelength form of the dispersion arises due to the long range nature of the Coulomb 
interaction and is thus independent of whether the superconductor is chiral or not. 
In analogy with the usual definition of the plasma frequency, we can define two different plasma frequencies 
corresponding to oscillations in the a-b plane , $\omega_{ab} ^p $ for $ \bm{k} = (k_x, k_y, 0) $ and along the 
z-direction, $\omega_c^p  $ for $\bm{k} = (0,0,k_z) $: 
\begin{eqnarray}
\omega_{ab}^p = \frac{4\pi \rho e^2}{m} \,\,;\,\,\omega_c^p = \frac{4\pi \rho e^2}{m_z}. 
\end{eqnarray}
 It follows from Eq.(\ref{current}), that the current of the collective mode for a superconductor has a transverse 
component $\frac{\cxy \omega}{e} (\hat{\bm{z}} \times \bm{k}) 
(\Phi/2) + \cxy (\bm{\hat{z} \times \bm{\nabla}} A_0)$
\footnote{We note that a neutral chiral superfluid would likewise
have a transverse component to the mass current, since the term 
 $\frac{\cxy \omega}{e} (\hat{\bm{z}} \times \bm{k}) 
(\Phi/2)$ persists in this case, though the term $\cxy (\bm{\hat{z} \times \bm{\nabla}} A_0) $ would be absent. }.
 This is  in addition to any transverse components which may arise 
from mass anisotropy.  To differentiate this
mixing of longitudinal and transverse degrees of freedom from that which arises purely due to mass 
anisotropy, one can consider the  2D limit $m_z \rightarrow \infty $ for $\bm{k}$ vectors lying 
in the x-y plane. It is clear that a chiral 2 d  superconductor, even one with Galilean invariance \footnote{The condition of Galilean invariance can be relaxed to that of threefold or higher rotational  symmetry as before.} will have 
collective modes which have a transverse component. Due to the chiral nature of the order, the 
collective modes are no longer purely longitudinal.

\section{III. Response of a chiral 3D superconductor} 
 The current in Eq. (\ref{current}) can be written as 
\bea
 \bm{j} = \bm{j}^{0} + \bm{j}^{cs},
\eea
where $ \bm{j}^{cs} $ contains all the terms proportional to $\cxy$ and no other terms and $\bm{j}^{0}$ is the current 
density for $\cxy = 0 $. 
 For the remainder of the paper, our focus shall be on the anomalous part of the current.

 From Eqs. (\ref{current}) and (\ref{5}), the  anomalous part of the current response can be written in terms of the external electric and magnetic fields as  
\bea
 \bm{j}^{cs} &=& { \cxy  ( \hat{\bm{z}} \times \bm{k})c_s ^2  (\bm{\tilde{k}\cdot E}) \over(\omega^2 - 
 c_s^2\bm{k}\cdot \bm{\tilde{k}})}   + { i c_s ^2 \tilde{\bm{k}} \cxy i \omega B_z  \over (\omega^2 - c_s^2\bm{\bm{k}\cdot\tilde{k}} )} +\nonumber \\
& & + {(\cxy)^2   \omega (\hat{\bm{z}} \times \bm{k} )i\omega B_z \over {e\rho_s \over mc_s^2} (\omega^2 -  c_s^2 \bm{k}\cdot\bm{\tilde{k}})}.   \label{jcs}
\eea

 We can express the current entirely in terms of the electric field, by writing 
$ i\omega B_z = i (\bm{k \times E})\cdot\bm{\hat{z}} $ and, hence, obtain the anomalous 
part of the conductivity tensor. The anomalous conductivity can be split up into two parts,  an antisymmetric part coming from the first term and a symmetric part which comes from the second term :
\bea
 \sigma^a _{im} = {\cxy c_s^2 \over (\omega^2 - c_s^2\bm{k}\cdot\bm{\tilde{k}})} ( \epsilon_{3li} k_l \tilde{k}_m - 
 \epsilon_{3lm}\tilde{k}_i k_l ) \\
\sigma^s _{im} =  {\cxy^2 c_s^2 \omega m\over e\rho_s(\omega^2 - c_s^2\bm{k}\cdot\bm{\tilde{k}})} i 
\left(\delta_{ln}\delta_{im} - \delta_{lm}\delta_{ni}\right)k_l k_n .
\eea 

 The  in-plane Hall conductivity, $\sigma_{xy}$ is 
 \bea
 \sigma_{xy} = {-\cxy c_s^2 k_{\parallel} ^2 \over (\omega^2 - c_s^2\bm{k}\cdot\bm{\tilde{k}})} ,
 \eea
 which reduces to the two dimensional expression, Eq. (\ref{4}) scaled by a factor of $\alpha$ in the limit $m_z\rightarrow \infty$.

 The charge density can also be expressed as follows : 
\bea
  \delta \rho = - {\cxy c_s ^2 \bm{k}\cdot\bm{\tilde{k}} B_z \over (\omega^2 - c_s^2\bm{k}\cdot\bm{\tilde{k}})} + {e^2 \rho_s 
  \over m} {i \tilde{\bm{k}}\cdot\bm{E} 
\over (\omega^2 - c_s^2\bm{k}\cdot\bm{\tilde{k}})}, \label{charge}
\eea
and the internal field generated is
\begin{eqnarray}
\bm{E}_{int} &= &- i A^0 _{int} \bm{k} = {-i 4\pi\delta\rho \over
\epsilon k^2}\bm{k}\\
 &=& {i 4\pi\tilde{c}_{xy}c_s ^2 \bm{k\cdot\tilde{k}} B_z \bm{k}\over
\epsilon k ^2(\omega ^2 - c_s^2 \bm{k\cdot\tilde{k}})} +
{\omega_p ^2 (\bm{\tilde{k}\cdot E}) \bm{k} \over \epsilon k^2(\omega ^2 - c_s^2
\bm{k\cdot\tilde{k}})} .
\end{eqnarray}
The vector $\bm{E}$ on the right hand side of the equation represents the total electric field.  To find the 
 dispersion of the collective modes, we can set $ B_z = 0 $ and
$ \bm{E} = \bm{E}_{int} $ and recover the same dispersion relation that we found
previously.

 The effective action written above is valid at zero temperature in the low energy limit where $v_f k, \omega \ll \Delta $. At higher frequencies and non zero temperatures, the effective action in momentum space is more conveniently written down in momentum space. The terms $\cxy, c_s,\rho$ acquire a frequency and temperature dependence. In addition, there is an extra term in the effective action : 
\bea
 S = \int d^3 k d\omega \sum_{l} c_{l0}(\bm{k},\omega) A_l(-\bm{k},-\omega)A_0 (\bm{k},\omega).
\eea

This term can be neglected in the long wavelength and low frequency limit because it is proportional to both $\bm{k}$ and $\omega$ in this limit. At high frequencies and long wavelengths, the linear dependence on $\bm{k}$ remains, however the frequency dependence changes and this term becomes comparable to other terms in the effective action. 
This term results in the presence 
of the following additional terms in the current and density response:  
  \bea
  \bm{j_l}^{'}(\bm{k},\omega) = c_{l0}(\bm{k},\omega)(A_0 - i{\omega\over e} \Phi(\bm{k},\omega)) \label{currentsym}\\
  \bm{\rho}^{'}(\bm{k},\omega) = - c_{l0}(\bm{k},\omega) (A_l - i{k_l\over e}\Phi(\bm{k},\omega)) \label{densitysym}
  \eea
  
  The response of the collective phase variable becomes :
\bea
  {\Phi/2} = \frac{ \tilde{c}_{xy} i\omega B_z + {e^2 \rho_s \over m} (i \tilde{\bm{k}}\cdot \bm{A} -
   {i\omega A_0\over  c_s ^2 }) }
  {{e\rho_s \over m c_s^2}(\omega^2 - c_s ^2\bm{k\cdot \tilde{k}})+ 2{ c_{l0}k_l\omega  \over e}}  \nonumber \\
 - \frac{ic_{l0}(k_l A_0 + A_l \omega)}{{e\rho_s \over m c_s^2}(\omega^2 - c_s ^2\bm{k\cdot \tilde{k}})+ 2{ c_{l0}k_l\omega  \over e}}    \label{phitot}
 \eea
  where a summation over repeated indices is implied and the dependence of $\rho, 
  \cxy , c_s $ on $\omega, T$ and of $c_{l0}$ on $(\bm{k},\omega, T)$ have been 
  suppressed.

 The full current and density response and the dispersion of the collective modes also changes and can easily be deduced  from Eqs. (\ref{jcs}), (\ref{charge}) and (\ref{currentsym})-(\ref{phitot}). Here we only write down the expressions for the anomalous charge density and the 
 antisymmetric part of the 
anomalous Hall conductance 
\bea
\delta\rho^{cs} = \cxy \frac{B_z(c_{lo}k_l \omega - e^2 {\rho_s \over m} \bm{k}\cdot\bm{\tilde{k}})}{{e^2\rho_s \over m c_s^2}(\omega^2 - c_s ^2\bm{k\cdot\tilde{k}})+ 2 {c_{l0}k_l\omega}}
\eea
\bea
\sigma_{im} = \cxy \left[ \frac{\epsilon_{3li}k_l (\tilde{k}_m{ e^2 \rho_s \over m} - c_{mo}\omega) } {{e^2\rho_s \over m c_s^2}(\omega^2 - c_s ^2\bm{k\cdot\tilde{k}})+ 2 {c_{l0}k_l\omega}}  \nonumber  \right.\\\label{sigmaim}  
\left.- \frac{\epsilon_{3lm}k_l (\tilde{k}_i {e^2 \rho_s\over m} - c_{i0}\omega) }{{e^2\rho_s \over m c_s^2}(\omega^2 - c_s ^2\bm{k\cdot\tilde{k}})+ 2 {c_{l0}k_l\omega}}     \right]
\eea 
 Using  Eq. (\ref{sigmaim}) and the results from the appendix, in the 
long wavelength and high frequency limit and the 2D limit of $ m_z\rightarrow \infty$ we deduce that 
\bea
 \sigma_{xy}  = \cxy   \left(- {k_{\parallel} ^2 p_f ^2 \over 2 m^2 \omega^2}\right)  \label{sigmaxy}
\eea
 where $p_f $ is the Fermi momentum. 

 \section{IV. Kerr Effect and Discussion}

One experiment which directly probes the anomalous ac Hall conductivity considered here is the polar Kerr 
effect \cite{xia:167002}.  In this experiment, linearly polarized light beam which is normally 
incident on the superconducting planes is reflected back as elliptically polarized light.  
The polar Kerr angle, which measures the degree of rotation of the polarization, is an indicator 
of the extent of time reversal symmetry breaking \cite{landau,white}. 
 If the c-axis of strontium ruthenate is 
chosen to be the z-axis, the 
Kerr angle is proportional to the real or imaginary part of 
$\sigma_{xy}$ depending on whether the real part of the refractive 
index is much smaller or larger than 
the complex part (see Appendix).
 
Connecting the calculations presented here to the Kerr effect experiments
reported in Ref. \cite{xia:167002} is problematic for two reasons.  First,
to a very good approximation, the experiment is done under conditions of 
light normally incident on the superconducting surface, i.e. $ \bm{k } =  (0,0,k)$, 
for which the Hall conductivity calculated above vanishes.  Second, the
frequency of the incident light, $\omega\simeq$0.8 eV, is very large compared 
to the superconducting gap of strontium ruthenate, $\Delta\simeq $0.23 meV, although probably 
not large enough to create transitions to higher lying energy bands \cite{oguchi}.
While the pairing interaction for strontium ruthenate is not known, this large
probing frequency is likely to be beyond the pairing cutoff used in BCS theory.  As
shown in the Appendix, the pairing cutoff enters the coefficient, $c_{xy}$, and the
ac Hall conductivity is substantially reduced for frequencies above this cutoff.  
Keeping these two caveats in mind, one can crudely estimate the predicted Kerr angle
if one simply assumes the experiment is probing at a frequency below the superconducting
pairing cutoff and at a finite in-plane wavevector introduced by the finite size
of the laser beam incident on the surface.  In 
this case, taking the complex refractive index to be $1.72i$ \footnote {This value is obtained by 
using $\epsilon = 10 $ and $\omega_p = 2.88 eV$  (based on optical data in the normal 
state \cite{katsufuji}) in  $ (n +i \kappa )^2 = \epsilon - \omega_p ^2/\omega^2$. 
It is desirable to obtain  optical data in the superconducting state.},   
one finds using Eqs. (\ref{sigmaxy}),(\ref{37}) and (\ref{Kerr2})  that the Kerr angle is of the order of $\sim 10^{-17} $ 
radians \footnote{We have used $k_{\parallel}\approx {2\pi\over l}$, where $l = 25\mu$m is the 
beam diameter and $k_f /m = 5.5 \times 10^{4}$ms$^{-1}$ corresponding to the gamma band \cite{mackenzie}.}, or roughly nine orders of magnitude smaller than the observed 
value. 

From Eqs. (\ref{Kerr}) and (\ref{Kerr2}), we see that the Kerr angle is greatly enhanced in 
the region where $ n+i\kappa \sim 0$, which corresponds to $\omega \sim \omega_p/\sqrt{\epsilon}$.
In this case, the Kerr angle should be determined using Eq. (\ref{35}).  However, for the
predicted Kerr angle to be of the order seen in experiments, the probing frequency would 
need to be within $\sim 10^{-6}\%$ of $\omega_p/\sqrt{\epsilon}$.  While disorder and lifetime
effects can be expected to broaden this window, the resonance is sufficiently sharp and the
enhancement required sufficiently large that this is very unlikely to explain the experiments.

Disorder can have a substantial effect on the conductivity, since it relaxes the
constraints imposed by Galilean invariance.  Consequently, disorder-induced terms can
contribute significantly to the conductivity tensor at finite frequency and zero wave vector, provided
the scattering rate is not too small relative to the superconducting gap.  The superconducting
transition temperature of Sr$_2$RuO$_4$ is very sensitive to disorder and samples exhibiting
the maximum $T_c$ of 1.5 K are believed to be in the clean limit.  Nevertheless, $\tau$
is still estimated to be $\sim 10^{-11} $s \cite{xia:167002} which could substantially
alter the results presented here, as well as earlier calculations which investigated some of
the quasiparticle and collective-mode contributions to the Kerr effect \cite{yip,joynt}. 
However, given the high probing frequency, the concern about being above
the pairing frequency cutoff remains.

More generally, our results suggest that it would be most interesting to study the Hall
conductivity or Kerr angle of Sr$_2$RuO$_4$ at lower frequencies where the signatures of
the chiral superconducting order are most pronounced.  Of course, it would also be of 
great interest to directly study the response at finite wave vector, for example,
by superimposing a grating on the sample, although it is always difficult to achieve 
wave vectors of sufficient size. 

{\it{Note added:}} While at KITP in December 2007, we learned that another group, Roman Lutchyn,
Pavel Nagornykh and Victor Yakovenko, had achieved similar results which were in substantial
agreement with ours, using a somewhat different approach.  We thank them for sending us a 
copy of their manuscript before posting it on the arxiv \cite{lutchyn}.  

\subsection{Acknowledgments}
 This work was supported by the Natural Sciences and Engineering Research Council of Canada (NSERC)
and by the Canadian Institute for Advanced Research. We would like to thank the Kavli Institute for Theoretical Physics, Santa 
Barbara (supported by the National Science Foundation under Grant No.PHY05-51164 ) where a part of this work 
was completed for its hospitality during the mini-program ``Sr$_2$RuO$_4$ and Chiral p-wave 
superconductivity''. We would also like to thank John Berlinsky, R Brout, Aharon Kapitulnik, Roman Lutchyn
Vladimir Mineev, Chetan Nayak, Jim Sauls,  
Michael Stone, and Victor Yakovenko for useful discussions.

\section{Appendix}
\subsection{A. High frequency response}
The parameters $c_{l0}, \cxy, c_s $ at high frequency can be calculated using linear response theory. For simplicity, we consider the $T=0 $ case. 
The expressions for 
the finite temperature coefficients can be calculated in an analogous manner \cite{arseev, yakovenko}. 
\def \omd {(\omega^2+ i\delta)^2 - (2\epsilon)^2}
\def \intdp {\int\frac{d^3 p}{(2\pi)^3}}
\bea
 {\rho_s \over mc_s^2} = -\int\frac{d^3 p}{(2\pi)^3} {4\Delta^2 \over \epsilon} \left(\frac{1}{\omd}\right) \label{soundvel}\\
 c_{l0} = e^2 \omega k_l \intdp \frac{(v_l )^2 \Delta^2 }{\epsilon^3}\left(\frac{1}{\omd}\right) \label{cl0}\\
 \cxy = -2 e^2 \intdp \frac{v_x f(p)}{\epsilon} \frac{1}{ (\omd)}   \label{cxy}
\eea
 where 
\bea
 f(p) = \lim_{q_y\rightarrow 0}\frac{Im\left(\Delta(p_x, p_y + {q_y\over 2})\Delta^{\dagger}(p_x, p_y-{q_y \over 2})\right)} {q_y}, 
\eea
$
\epsilon (p) = \sqrt{\xi(p)^2 + \Delta(p)^2} $
 and $\xi(p)$ are the single particle energies of the electronic system and 
$v_l = \frac{\partial \xi(p)}{\partial p_l}$.

To evaluate these integrals, we take the limit $m_z\rightarrow \infty$ and use the free particle dispersion in 2D: $\xi(p) = p^2/2m - \epsilon_f$ 
where
$\epsilon_f$ is the Fermi energy. We take the gap function 
to be 
\bea
\Delta(p) = \left\{ 
\begin{array}{cl} 
\Delta_0 (p_x +i p_y)/p & \text{if}\,\,|\xi(p)| < \omega_D \\
& \\ 
0 & \text{if} \,\, |\xi(p)|> \omega_D
\end{array}
\right.
\eea 
 where $\omega_D \gg \Delta$ is a BCS cutoff. 
 Then $\cxy$ reduces to
\bea
\cxy &=& {e^2\over 8\pi a} \int_{-\omega_D} ^{\omega_D} \frac{dx}{\sqrt{1+x^2}(1+x^2-({\omega/2\Delta_0})^2)} 
\eea
\bea
  \cxy={ e^2\over 4\pi a}\left\{ 
  \begin{array}{cl}
    \frac{\sin^{-1}(\alpha)}{\alpha\sqrt{1-\alpha^2}}& \,\,\text{for} \,\,\omega < 2 \Delta \\
   &  \\
{i\pi\over 2\alpha^2} - {1\over \alpha^2}\ln({\omega\over\Delta})&\,\,\text{for}\,\, 2\Delta\ll\omega <\omega_D \\
 &\\
 -\frac{1}{\alpha^2}\ln({\omega_D\over\Delta}) &\,\,\text{for}\,\, \omega_D\ll\omega
 \end{array}\right.   \label{37}
\eea
where $\alpha= {\omega \over 2\Delta_0}$ and a is the interlayer spacing. This reduces in the limit  $\omega_D \rightarrow \infty $ to the 
expressions obtained in Ref. \cite{yakovenko} scaled by a factor of $1/a$.

While the BCS cutoff is a crude approximation to the energy dependence of any realistic
pairing potential, it is used here to highlight the fact that the effects of the Chern-Simons-like
term are only effective close to the Fermi energy.  As pointed out by Yakovenko \cite{yakovenko},
$\cxy$ corresponds to the excitation of two BCS quasiparticles (or quasiholes).  Such a term typically
carries the usual coherence factor, but $\cxy$ only carries the piece containing the chiral signature,
i.e. $\Delta(p)$, which vanishes at energies above the pairing cutoff.

To obtain the in-plane Hall conductivity $\sigma_{xy}$ in the long wavelength limit at high frequencies, we also need to calculate
\bea
\left( \frac{ m c_S ^2 (-(k_x ^2 + k_y ^2){e^2 \rho_s \over m} + \omega(c_{y0} k_y + c_{x0} k_x))}{e^2 \rho_s \omega^2 } \right)
\eea
 Using Eqs. \ref{soundvel},\ref{cl0} and \ref{cxy} ,this quantity reduces at high frequencies in the two dimensional limit of $m_z 
\rightarrow 0 $ to 
$\left(- {k_{\parallel} ^2 p_f ^2 \over 2 m^2 \omega^2} \right) $ \footnote{We thank Roman Lutchyn for a useful discussion on this
 point. Also see Ref. \cite{lutchyn}.}. 

  \subsection{B. Kerr Angle}
  Let $ n+i\kappa$ be the complex refractive index given by $ (n + i \kappa)^2 = \epsilon + {i4\pi \sigma_{xx} \over \omega}$ 
and $ n_{\pm} + i \kappa_{\pm} $ be the complex index of refraction for right and left circularly polarized light. Then \cite{white,landau},  
\bea
 (n_{\pm} + i\kappa_{\pm})^2 = 1 + i4\pi\sigma_{\pm}/\omega
\eea
 where $\sigma_{\pm} = \sigma_{xx}\pm i\sigma_{xy}$.
 The Kerr angle is given by : 
\bea
\theta_{\kappa} = -{1\over2}\left( -tan^{-1}({\kappa_+ \over 1-n_{+}}) + tan^{-1}({\kappa_{-} \over 1+ n_{-}})  \right.\nonumber\\
\left.  -tan^{-1}({\kappa_+ \over 1 + n_{+}}) + tan^{-1} ({\kappa_- \over 1 -n_{-}})\right) \label{35}
\eea
 When $ n \gg \kappa$, then the Kerr angle is given by the formula 
\bea
\theta_{K} = \frac{4\pi \sigma^{''}_{xy}} {n (n^2 -1) \omega} \label{Kerr}
\eea and when $ \kappa \gg n $, then
\bea
 \theta_{K} = \frac{4\pi \sigma_{xy} ^{'}}{\omega \kappa ^3} \label{Kerr2}
\eea

In the clean limit, when $\omega $ is close to $\omega_p/\sqrt{\epsilon}$, Eq. (\ref{35}) should be used to 
calculate the Kerr angle, while far from the resonance, Eqs. (\ref{Kerr}) and (\ref{Kerr2}) should be used when $\omega > \omega_p/\sqrt{\epsilon}$ and $ \omega < \omega_p/\sqrt{\epsilon} $ respectively.

\end{document}